\documentclass[twocolumn,showpacs,preprintnumbers]{revtex4}
\usepackage{amsmath}
\usepackage{amssymb}
\usepackage{graphicx}
\usepackage{bm}
\begin{document}
\title{Comparison of dynamical multifragmentation models}
\author{J. Rizzo$^{a,b}$, M. Colonna$^{a,b}$, A. Ono$^{c}$} 
\affiliation{
        $^a$LNS-INFN, I-95123, Catania, Italy\\
	$^b$Physics and Astronomy Dept. University of Catania, Italy\\
        $^c$Department of Physics, Tohoku University, Sendai 980-8578,
      Japan}

\begin{abstract}
  Multifragmentation scenarios, as predicted by antisymmetrized
  molecular dynamics (AMD) or momentum-dependent stochastic mean-field
  (BGBD) calculations are compared.  While in the BGBD case fragment
  emission is clearly linked to the spinodal decomposition mechanism,
  i.e. to mean-field instabilities, in AMD many-body correlations have
  a stronger impact on the fragmentation dynamics and clusters start
  to appear at earlier times.  As a consequence, fragments are formed
  on shorter time scales in AMD, on about equal footing of light
  particle pre-equilibrium emission.  Conversely, in BGBD
  pre-equilibrium and fragment emissions happen on different time
  scales and are related to different mechanisms.

\end{abstract}
\pacs{24.10.Pa; 25.70.Pq}
\maketitle

\section{Introduction}
During the last decade, multifragmentation, i.e. the break-up of excited nuclear
systems into many pieces, has been extensively investigated in heavy ion collisions (HIC)
around Fermi energies, both from the experimental and theoretical points of 
view \cite{Frankland,Moretto,Dagostino,Chomaz,EPJA,Ono,rep,Xu00,Ger04}.
In particular, the study of the mechanism responsible for fragment
production has driven much attention, especially in relation to the possibility
to observe a liquid-gas phase transition in nuclei \cite{Frankland,Moretto,Dagostino,Chomaz,EPJA}.
Many efforts have been devoted to the characterization of the 
properties of the fragmenting source,
such as temperature and density, to determine its location inside the
nuclear matter phase diagram. 

Due to compression and/or thermal effects,
the composite systems formed in HIC
may reach low density values, attaining the co-existence zone of the
nuclear matter phase diagram. 
For instance, an excited system that expands under
the conditions of thermal equilibrium could perform a phase transition 
staying close to the liquid branch of the co-existence line \cite{mononuclear,Moretto}.
However, due to the Coulomb instabilities, the
limiting temperature, that a nucleus can sustain as a compact configuration, 
may be lower than the critical temperature \cite{Nato}.
In this situation, the system is
brought inside the co-existence zone of the nuclear matter phase diagram and
undergoes a spontaneous phase separation, breaking up into several fragments \cite{rep,Chomaz}.
 
It is generally believed that in central HIC at Fermi energies, the composite matter 
can be compressed up to twice the normal density value
(as revealed for instance by the emission of energetic particles through hard two-body
scattering \cite{sap}), and then the system 
expands and breaks up into many pieces \cite{rep,EPJA,Ono}.  
One of the most challenging and still open questions is the understanding of
the fragmentation mechanism along this path. 
The decompression following the initial 
collisional shock should be strong enough to push the system  inside 
the unstable region of the phase diagram and fragments could be 
formed due to the development of mean-field spinodal instabilities 
\cite{rep,EPJA}. 
However, already in the high density phase nucleon correlations 
are expected to be rather large, due to the huge amount of
two-body nucleon-nucleon collisions.  Hence some memory of these high density correlations
could be kept along the fragment formation process. 

According to classical molecular dynamics \cite{Dorso,Campi} or lattice-gas  
calculations \cite{Francesca}, 
self-bound clusters are observed, even at equilibrium, in high density
systems. 
This has recently suggested an interpretation of the multifragmentation
phenomenon in terms of a sudden explosion of the system, where 
pre-fragments start to appear already in the high density phase and
they subsequently fly apart from each other due to the strong Coulomb
repulsion \cite{bigbang}.  
However, one should keep in mind that this clustering
effect revealed in the high density phase could be much stronger 
in classical systems than in nuclear matter, which is a Fermi liquid. 
For instance, mean-field calculations of the response of nuclear matter
to the presence of density fluctuations show that  
correlations constructed in the high density phase are damped. Indeed,
according to a mean-field description, 
it is not energetically convenient for a Fermi system at density larger than the
normal value to develop these high density bumps \cite{rep}. 

\begin{figure*}[t]
\includegraphics[width=18.cm]{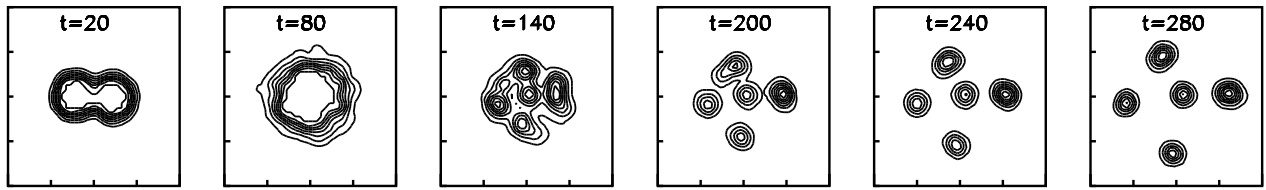}
\caption{Contour plots of the density projected on the reaction plane
  calculated with BGBD for the central reaction $^{112}$Sn +
  $^{112}$Sn at 50 MeV/nucleon, at several times (fm/$c$). 
The size of each box is 40 fm.}
\label{contour_BGBD}
\end{figure*}
\begin{figure*}[t]
\includegraphics[width=18.cm]{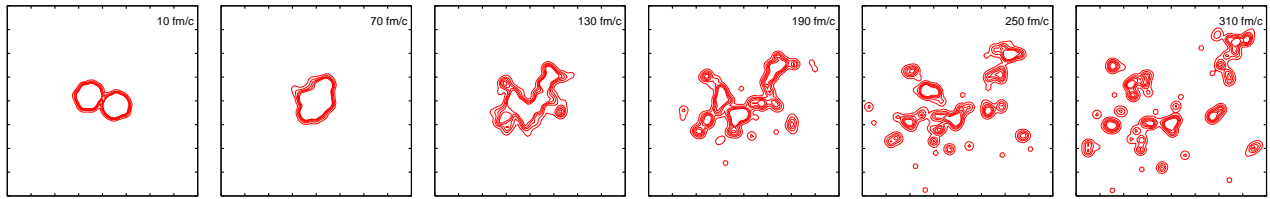}
\caption{The same as in Fig.\ \ref{contour_BGBD} but calculated with
  AMD. The size of each box is 80 fm.}
\label{contour_AMD}
\end{figure*}

However, if a large amount of fluctuations is present in the high density stage of the reaction,
and the system expands rather quickly,  
these correlations could survive anyway and play an important role in determining the
properties of the formed fragments. 
On the other hand, if fluctuations are not so large, 
these structures could be re-organized again by the low
density unstable mean-field.

Hence it is clear that the fragmentation mechanism is very sensitive to the 
delicate balance between many-body correlations, 
mean-field effects and the time scale imposed 
by the reaction dynamics, especially in the Fermi energy domain, where 
one- and two-body effects are equally important.
Therefore, 
depending on the way one treats the different effects, one could
expect a different outcome from the available theoretical models
describing multifragmentation. 
In this paper we undertake this kind of investigation, by comparing the
results given by two fragmentation models: the stochastic mean-field, 
including momentum dependence (BGBD) \cite{Joseph_model} and the 
antisymmetrized molecular dynamics (AMD) models \cite{Ono}.
Both have proved to give a good reproduction of 
some aspects of multifragmentation data \cite{ONOj,Frankland1}. 
We analyze central reactions, where we expect to see larger dynamical 
(compression-expansion) effects
on fragment formation. 
The present study will allow us to get a deeper insight into the reaction
mechanism, in connection with the ingredients of the two models.
This should be reflected into the properties
of the obtained primary fragments and eventually on measured observables.
In this way one can also try  to identify the experimental quantities 
that are more sensitive to the fragmentation scenario.
The paper is organized as it follows:
In section II we will give an outlook of the main ingredients of both models. 
In section III we study the fragmentation path, as given by the two models, 
in the case of a central reaction, $^{112}$Sn + $^{112}$Sn  at 50 MeV/nucleon
\cite{Xu00}.
Conclusions are drawn in section IV. 

\section{Ingredients of the models}
Two different kinds of microscopic approaches have been proposed and applied
to study heavy ion reaction mechanisms, i.e. to describe the dynamics of
nuclear many-body systems. 
One is the class of molecular dynamics models \cite{md,feld,ONOab,Pap,FFMD} while the other
kind is represented by stochastic mean-field approaches \cite{Ayik,Randrup,
rep}.

In the mean-field class of descriptions, the dynamical state of
the nuclear system is characterized by the reduced one-body density
in phase space, $f({\bf r},{\bf p},t)$, the classical analogue of the
Wigner transform of the one-particle density matrix.
At low energies, the time evolution of the one-body density is governed by the
Vlasov equation, which can be regarded as the semi-classical approximation 
to the time-dependent Hartree-Fock theory. The residual direct collisions
between the constituent nucleons are incorporated by means of a Pauli-blocked
collision integral, leading to the so-called Boltzmann-Uehling-Uhlenbeck (BUU) 
or Boltzmann-Nordheim-Vlasov (BNV) approaches \cite{Bert,Aldo}. 
The stochastic extension of the transport treatment for the one-particle
density is obtained by introducing a stochastic term representing the 
fluctuating part of the collision integral \cite{Ayik}, in close analogy
with the Langevin equation for a brownian motion.
 
Molecular dynamics models usually assume a fixed Gaussian shape for 
the single particle wave functions. The many-body state is represented 
by a simple product wave function, with or without antisymmetrization.
In this way, though single particle wave functions are supposed to be
independent (mean-field approximation), the use of localized wave packets
induces many-body correlations both in mean-field propagation and 
hard two body scattering (collision integral), that is treated stochastically. 
From the point of view of stochastic mean-field models, the philosophy of
molecular dynamics would be to introduce a special kind of fluctuation
by stochastically localizing the single particle wave functions.

Below we will give the ingredients of the two models, that can be seen as
representative of each class, that we consider in the present work.

\subsection{BGBD}

The stochastic mean-field model considered here is a
semi-classical non-relativistic transport approach, of 
BNV
type, see \cite{Alfio,BaranPR410}, using an isospin and momentum
dependent effective interaction. The latter is  
derived via an asymmetric extension of the Gale-Bertsch-DasGupta (GBD) force
 \cite{GalePRC41,GrecoPRC59}.

The energy density  can be parametrized
as follows (see also \cite{Isospin01},\cite{Bombiso}):
\begin{equation}
\varepsilon=\varepsilon_{\text{kin}}+\varepsilon(A',A'')
+\varepsilon(B',B'')+\varepsilon(C',C'')
\label{edensmd1}
\end{equation}
where $\varepsilon_{\text{kin}}$ is the usual kinetic energy density and
\begin{gather}
\varepsilon(A',A'')=(A'+A''\beta^2)\frac{\rho^2}{\rho_0}
 \nonumber \\
\varepsilon(B',B'')=(B'+B''\beta^2)\left ( \frac{\rho}{\rho_0}
\right )^{\sigma}\rho
 \nonumber \\
\varepsilon(C',C'')=C'(\mathcal{I}_{NN}+\mathcal{I}_{PP})
+C''\mathcal{I}_{NP}
\label{edensmd2}
\end{gather}
The variable $\beta=(N-Z)/A$ defines the isospin content of the system,
given the number of neutrons $(N)$, protons $(Z)$, and the total mass $A=N+Z$;
the quantity $\rho_0$ is the normal density of nuclear matter.
The momentum dependence is contained in the $\mathcal{I}_{\tau \tau'}$
terms, which indicate integrals of the form:

$$\mathcal{I}_{\tau \tau'}=\int d \vec{p} \; d \vec{p}\,' 
f_{\tau}(\vec{r},\vec{p})
 f_{\tau'}(\vec{r},\vec{p}\,') g(\vec{p},\vec{p}\,')$$
where $g(\vec{p},\vec{p}\,')=1/(1 + (\vec{p}-\vec{p}\,')^2/\lambda)$
and $f_{\tau}$ represents the one-body distribution function of neutrons or
protons. $\lambda$ is a constant, that is taken equal to $(1.5~ k_F)^2$, 
being $k_F$ the Fermi momentum at normal density. 
This choice of the function $g(\vec{p},\vec{p}\,')$ 
gives a similar behaviour with respect to the Gogny effective interaction
used in the AMD simulations (see next sub-section).
We use a soft equation of state for symmetric nuclear matter 
(compressibility modulus $K_{\text{NM}}(\rho_0)=215$ MeV).
In this frame
we can easily adjust the parameters in order to fix the density
dependence of effective mass and symmetry energy.  


So we describe the time evolution of the system in terms of the one-body
distribution function $f_\tau({\bf r}, {\bf p}, t)$, as ruled by the 
nuclear mean-field (plus Coulomb interaction for protons) and hard
two-body scattering, according to the so-called Boltzmann-Langevin equation \cite{Ayik,Randrup}:
\begin{equation}
{{df_\tau}\over{dt}} = {{\partial f_\tau}\over{\partial t}} 
+ \{f_\tau,H\} = I_{\tau}[f] 
+ \delta I_\tau[f],
\label{BL}
\end{equation}
 where 
 $H({\bf r},{\bf p},t)$ is the one-body Hamiltonian,
$I_\tau[f]$ is the average two-body collision integral and 
$\delta I_\tau[f]$ represents the stochastic source term \cite{Ayik,Randrup,rep}.
The test particle method is used to solve numerically Eq.\ (\ref{BL}). 
The free energy- and angle-dependent nucleon-nucleon cross section is used
in the collision integral.  
 Fluctuations are introduced within this mean-field
treatment, according to the approach presented in \cite{Alfio,Salvo}, i.e. by agitating 
the spatial density profile. 
Once local thermal equilibrium is reached, the density fluctuation amplitude, $\sigma_{\rho}$ is
evaluated by projecting on the coordinate space the kinetic equilibrium value
of a Fermi gas. Then,
in the cell of ${\bf r}$ space being considered, the density fluctuation
$\delta\rho$ is selected randomly according to the gaussian distribution
$\exp(-\delta\rho^2/2\sigma_{\rho}^2)$. This determines the variation of the
number of particles contained in the cell. A few left-over particles are 
randomly distributed again to ensure the conservation of mass. 
Momenta of all particles are finally slightly shifted to ensure momentum
and energy conservation.
Hence, while
the dynamical evolution of the system is still described in terms of the one-body distribution 
function,
this function experiences a stochastic evolution, in response to the action of a
fluctuation term essentially
determined by the degree of thermal agitation present in the system. 

According to this stochastic mean-field theory, the fragmentation process
is dominated by the growth of volume (spinodal) and surface instabilities encountered during the expansion phase 
of the considered excited systems \cite{frag_path}. 
Therefore density fluctuations provide the seeds of the
formation of fragments, whose characteristics are related to the properties of the most unstable
collective modes of the mean-field. In finite nuclei, 
several multipoles are excited 
with close probabilities.
Hence a large variety of fragment configurations may be obtained, due to the beating of the
several unstable modes \cite{rep,EPJA,frag_path}.
This description of the fragmentation path can explain several features, 
concerning also rather exclusive observables \cite{EPJA}, 
observed in experimental data at around 30 MeV/nucleon \cite{Frankland1}.




\subsection{AMD}

For the AMD \cite{ONOab,ONOj,ONO-ppnp} calculations presented here, we
use the same framework as in Ref.\ \cite{ONOj} which can reproduce the
fragment charge distribution of the central $\mathrm{Xe}+\mathrm{Sn}$
collisions at 50 MeV/nucleon.

In AMD, we employ the Slater determinant of Gaussian wave packets
\begin{equation}
\langle\mathbf{r}_1\ldots\mathbf{r}_A|\Phi(Z)\rangle \propto
\det_{ij} \biggl[ \exp\Bigl\{-\nu(\mathbf{r}_i-\mathbf{Z}_j/\sqrt{\nu})^2\Bigr\}
                               \chi_{\alpha_j}(i) \biggr],
\label{eq:AMDWaveFunction}
\end{equation}
where $\chi_{\alpha_i}$ are the spin-isospin states with
$\alpha_i=p\uparrow$, $p\downarrow$, $n\uparrow$, or $n\downarrow$.  Thus
the many-body state $|\Phi(Z)\rangle$ is parametrized by a set of
complex variables $Z\equiv\{{\mathbf{Z}}_i\}_{i=1,\ldots,A}$, where
$A$ is the number of nucleons in the system.  The width parameter
$\nu=(2.5\ \textrm{fm})^{-2}$ is treated as a constant parameter
common to all the wave packets.  If we ignore the antisymmetrization
effect, the real part of $\mathbf{Z}_i$ corresponds to the position
centroid and the imaginary part corresponds to the momentum centroid.
This choice of wave functions is suitable to describe fragmentation channels  
where each single particle wave function should be localized within a
fragment.

The dynamics of fragmentation is a highly complicated quantum
many-body problem in which huge number of fragmentation channels will
appear in the course of the evolution.  An AMD wave function [Eq.\
(\ref{eq:AMDWaveFunction})] is intended to describe one of the
channels rather than the total many-body state, and the emergence of
channels is represented approximately by some stochastic terms in the
equation of motion.

The stochastic equation of motion for the wave packet centroids $Z$
may be symbolically written as
\begin{equation}
\frac{d}{dt}\mathbf{Z}_i
=\{\mathbf{Z}_i,\mathcal{H}\}_\text{PB}
+\mbox{(NN coll)}
+\Delta\mathbf{Z}_i(t)
+\mu\,(\mathbf{Z}_i,\mathcal{H}').
\end{equation}
The first term $\{\mathbf{Z}_i,\mathcal{H}\}_\text{PB}$ is the
deterministic term derived from the time-dependent variational
principle with an assumed effective interaction such as the Gogny
interaction \cite{GOGNY}.  The second term represents the effect of
the stochastic two-nucleon collision process. 
The collisions are performed with the ``physical nucleon coordinates''
that take account of the antisymmetrization effects, and then the
Pauli blocking in the final state is automatically introduced
\cite{ONOab}.  The third term $\Delta\mathbf{Z}_i(t)$ is a stochastic
fluctuation term that has been introduced in order to respect the
change of the width and shape of the single particle distribution
\cite{ONOh,Ono,ONOj}.  In other words, the combination
$\{\mathbf{Z}_i,\mathcal{H}\}_\text{PB}+\Delta\mathbf{Z}_i(t)$
approximately reproduces the prediction by mean field theories 
for the ensemble-averaged single-particle
distribution, while each nucleon is localized in phase space for each
channel.  The term $\Delta\mathbf{Z}_i(t)$ is calculated practically
by solving the Vlasov equation 
with the same
effective interaction as for the term
$\{\mathbf{Z}_i,\mathcal{H}\}_\text{PB}$.  In the present version of
AMD \cite{ONOj}, the property of the fluctuation
$\Delta\mathbf{Z}_i(t)$ is chosen in such a way that the coherent
single particle motion in the mean field is respected for some time
interval until the nucleon collides another nucleon.  The last term
$\mu\,(\mathbf{Z}_i,\mathcal{H}')$ is a dissipation term related to
the fluctuation term $\Delta\mathbf{Z}_i(t)$.  The dissipation term is
necessary in order to restore the conservation of energy that is
violated by the fluctuation term.  The coefficient $\mu$ is given by
the condition of energy conservation. However, the form of this term
is somehow arbitrary.  We shift the variables $Z$ to the direction of
the gradient of the energy expectation value $\mathcal{H}$ under the
constraints of conserved quantities (the center-of-mass variables and
the total angular momentum) and global one-body quantities (monopole
and quadrupole moments in coordinate and momentum spaces).  A complete
formulation of AMD can be found in Refs.\ \cite{ONOj,ONO-ppnp}.

In the present work, we use the Gogny effective interaction
\cite{GOGNY} which corresponds to a soft equation of state of
symmetric nuclear matter with the incompressibility
$K_{\text{NM}}(\rho_0)=228$ MeV.  The mean field for this force has
a momentum dependence which is similar to that in the BGBD
calculation.  As the two-nucleon collision cross sections
($\sigma_{pp}=\sigma_{nn}$ and $\sigma_{pn}$), we use the energy- and
angle-dependent values in the free space with the maximum cut-off of
150 mb.  In order to avoid low energy spurious collisions, due to
Pauli blocking violation (caused by the finite number of test
particles), in the BGBD calculations a lower cut-off, 50 mb, has been
considered.  We have checked that this leads to approximately the same
degree of stopping in the two models.

\subsection{Some remarks}
We would like to stress here the main differences between the two
models described above. 

In the AMD version used here \cite{ONOj}, a special procedure is
adopted, that ensures a coherent single-particle motion in the
mean-field (including diffusion and shrinking effects of the nucleon wave
packet in phase space) until the considered nucleon collides another one. 
It can be demonstrated that this procedure reproduces
exactly the coherent time evolution of the Wigner function $f$ in the
case of a harmonic oscillator potential (or for free nucleons), though  
this correspondence is not exact in the general case.  Hence
one-body effects should be similarly treated in the two models.  
From this point of view, the present version of AMD is rather 
different from earlier molecular dynamics
formulations \cite{md,ONOab}, where the use of localized wave packets in the full dynamics
implies that one-body effects are not as precisely described as in
mean-field models. 

The most relevant difference between AMD and BGBD 
is related to the method followed to 
implement stochastic
two-body scattering.
In fact, in BGBD fluctuations are introduced by
agitating the one body density function, to account for the stochastic
part of the collision integral, according to the Boltzmann-Langevin
theory \cite{Ayik,rep}. In some sense, this would correspond to a
description of the system in terms of unrestricted fluctuating single
particle wave functions.  On the other side, in AMD fluctuations are
introduced by stochastically localizing the single particle wave
functions in phase space when a two-nucleon collision takes place.

It is difficult to discuss 
the general validity of the various  approximations adopted to solve 
the quantum many-body dynamics because it may depend    
on the particular reaction mechanism and energy
range under study.  
Here we will focus on the description of the multifragmentation
mechanism at 50 MeV/nucleon.
While it is well known that the results of early molecular dynamics models \cite{md,ONOab}
and standard mean-field approaches \cite{Aldo}
are rather different, 
the improved AMD and the stochastic BGBD can be considered as closer
approaches and 
it is interesting to investigate 
how their respective  predictions compare to each other 
in the case of fragmentation reactions.

\section{Analysis of the results}
Our aim is to investigate the fragmentation path in violent collisions
at intermediate energy, as predicted by the BGBD and the AMD models.
As discussed above, in the BGBD model one essentially follows the
evolution of the one-body density and the fragmentation mechanism is
mainly based on the amplification of its fluctuations.  In AMD nucleon
wave packets are propagated, from which, however, it is possible to
reconstruct the one-body density and related observables.

Hence our study of the reaction path will be performed by looking at
quantities connected to the one-body density and to its fluctuations.
We will investigate central collisions of the system $^{112}$Sn +
$^{112}$Sn at 50 MeV/nucleon.  An ensemble of 200 trajectories with $b
= 0.5$ fm has been collected with BGBD, while in AMD 20 events with $0 < b
< 1$ fm are considered.

\subsection{One-body observables}
To give a qualitative representation of the time evolution of the
system, density contour plots in the reaction plane, as obtained in
the two models, are shown in Figs.\ \ref{contour_BGBD} and
\ref{contour_AMD} at several time steps.  As one can see from the
figures, both models predict that the system is initially
compressed. Then expansion follows and several intermediate mass
fragments (IMF) appear.

\begin{figure}
\includegraphics[width=6.cm]{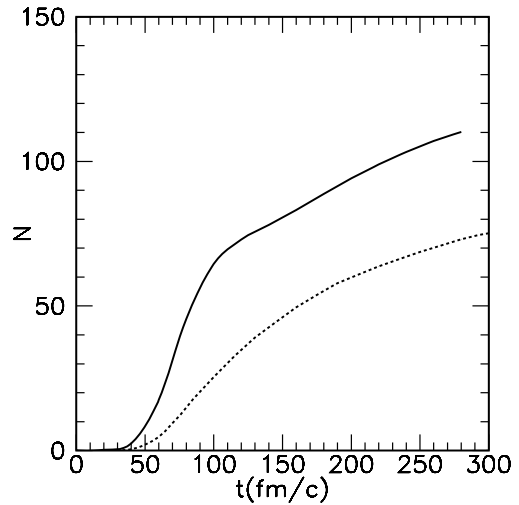}
\caption{Time evolution of the number of nucleons in the ``gas''
  phase, as obtained in the central reaction $^{112}$Sn + $^{112}$Sn
  at 50 MeV/nucleon. Full line refers to BGBD calculations, while
  dashes represent AMD calculations.}
\label{pre-equi}
\end{figure}

In BGBD, according to the value of the spatial density $\rho({\bf
  r})$, one can identify a ``gas'' phase ($\rho < \rho_0/3$),
associated with particles that leave rapidly the system
(pre-equilibrium emission) and/or are evaporated, and a ``liquid''
phase, where fragments belong to.  $\rho_0$ denotes the normal density
value.  As a matter of fact, we observe that the ``liquid''
essentially corresponds to particles with charge greater than 2.  In
AMD, the fragments with $A\le 4$ are regarded as in the gas part.  The
time evolution of the number of nucleons that are in the gas phase is
represented in Fig.\ \ref{pre-equi}.  It is possible to observe that
the number of particles that escape at early times from the interacting
nuclear matter, due to two-body scattering (pre-equilibrium effects),
is different in the two models.  In the BGBD case (full line) the
emission rate is larger with respect to the AMD
calculations. Moreover, there is a clear change of slope at around
$t\approx 100$ fm/$c$ where, as we will see in the following (see also Fig.1), the
nuclear system reaches low density values, fragments start to be
formed and nucleons are emitted due essentially to evaporative
processes. This is not so evident in the AMD case.  This result seems
to indicate already different emission mechanisms between the two
models.  In BGBD the production mechanism of pre-equilibrium light
particles and the following evolution of the system (fragmentation)
are somehow decoupled, while in the AMD case there are no easily
identifiable emitting sources and all fragments and light particles
seem to be emitted on about equal footing.
 
To closely follow the time evolution of the system and to better
characterize its fragmentation path, we have studied the behaviour of
the following observables: the radial density profile of the system
and the radial collective momentum as functions of time.  
The radial density at a given distance $r$ is obtained by averaging
the local density $\rho({\bf r})$ over the surface of a sphere of
radius $r$.  The radial collective momentum is the projection of the
collective momentum at the position ${\bf r}$ along the radial
direction, averaged over the surface of the sphere of radius $r$.
These quantities are further averaged over the event ensemble.

\begin{figure}
\includegraphics[width=7.cm]{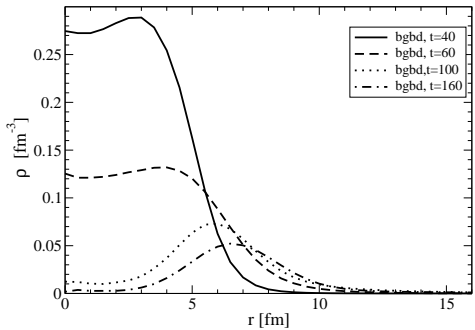}
\caption{Density profiles, at several times (in fm/c), as obtained in the BGBD case.}
\label{dens_BGBD}
\end{figure}

In the BGBD case, the behaviour of the radial density profile,
presented in Fig.\ \ref{dens_BGBD}, indicates that, after an initial
compression ($t = 40$ fm/$c$), the system expands and finally it gets
rather dilute, due to the occurrence of a monopole expansion,
generated by the compression.  As one can see from
Fig.\ \ref{contour_BGBD}, while the system expands, it breaks up into
many pieces.  
The matter
appears mostly concentrated within a given window of the radial
distance (see for instance the line at $t = 120$ fm/$c$), indicating
the formation of a bubble-like configuration, where fragments are
located.  Indeed the central region of the system is rapidly depleted.
\begin{figure}
\includegraphics[width=7.cm]{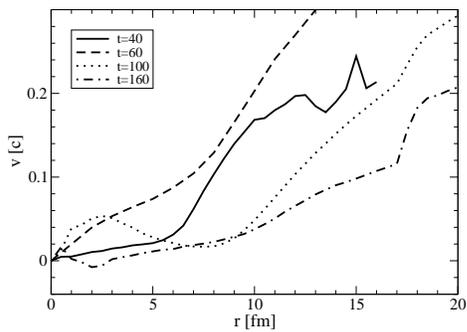}
\caption{Collective momentum profiles, in units of the nucleon free mass, 
at several times (in fm/c), as obtained in the BGBD case.
}
\label{vel_BGBD}
\end{figure} 



It is rather interesting to look at the profile of the collective
momentum (divided by the nucleon mass), $v(r)$, see Fig.\
\ref{vel_BGBD}. The time evolution of this quantity is largely
influenced by the occurrence of monopole compression and
expansion. Indeed, after the initial compression, the restoring force
generated by the mean-field pushes the system back to the normal
density value and the collective momentum increases (compare the
evolution from $t = 40$ to 60 fm/$c$).  It can be noticed that, at $t
= 60$ fm/$c$, in the region where the ``liquid'' is located, i.e.  for
$r<8$ - $9$ fm (cf. Fig.\ \ref{dens_BGBD}), the collective momentum
profile is almost self-similar, i.e.  the radial momentum is
proportional to the radial distance.  Then the system goes into the
expansion phase and the collective momentum decreases again.  
This deceleration indicates that the system is still rather 
homogeneous while it expands, reaching the low density (unstable) region.
Then  fragmentation can be associated with the occurrence
of spinodal decomposition \cite{rep}.
More specifically, the system slows down
due to the presence of a counter-streaming flow that develops from the
surface towards the interior, trying to recompact the system.  This
acts against the initial expansion, so the collective momentum profile
is modified and, in the region where the matter is located (cf. Fig.\
\ref{dens_BGBD}), the system is slowed down.  So actually when
fragments start to appear ($t \approx 100$ fm/$c$) their collective
momentum is not so large.  However, this is rather large in the
central part of the system ($r < 3$ fm at $t=60$ and 100 fm/$c$),
which is related to the rapid depletion of the density in this region.
\begin{figure}
\includegraphics[width=7.cm]{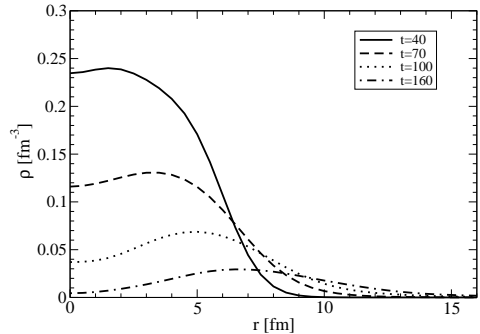}
\caption{Density profiles, at several times (in fm/c), as obtained in the AMD case.}
\label{dens_AMD}
\end{figure}

\begin{figure}
\includegraphics[width=7.cm]{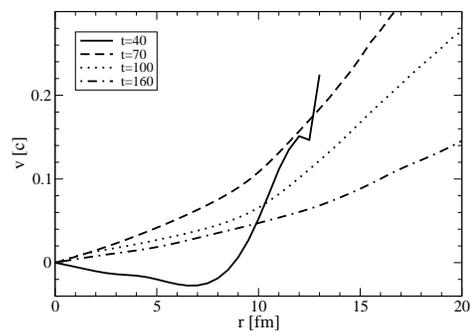}
\caption{Collective momentum profiles, in units of the nucleon free mass, at several times (in fm/c), as obtained in the AMD case.}
\label{vel_AMD}
\end{figure}  
One may also notice that, at large radial distance, where mostly
pre-equilibrium particles are located, the radial momentum exhibits a
different trend and is rapidly increasing with $r$.


In AMD calculations, the density profile in Fig.\ \ref{dens_AMD} shows
the time evolution of the compression and expansion which is
qualitatively similar to the BGBD case.  However, we notice that AMD
shows broader density distribution than BGBD as the system expands.
The bubble-like configuration is not as prominent as in BGBD.  The
analysis of the radial momentum profile (Fig.\ \ref{vel_AMD}) also
reveals that the expansion dynamics in AMD is different from the BGBD
case.  When the system goes from the compressed state towards the
normal density value ($t \approx 70$ fm/$c$), the
collective momentum increases less than in the BGBD case. This
suggests that the energy that was stored in compression goes into
larger (with respect to BGBD) thermal kinetic fluctuations, which is
consistent with the broader density profile observed in Fig.\
\ref{dens_AMD}.  
It should be noticed that
the broad density profile in the late stage does not
mean homogeneous dilute matter but corresponds to the situation in
which fragments are distributed widely in the space.  By the
comparison with the BGBD case, it seems that the system ceases to
behave as homogeneous matter already at around $t = 70$ fm/$c$, before
entering the low density (spinodal) region.  Then, contrarily to what
happens in the BGBD case, where the system is slowed down during the
expansion phase, after $t = 70$ fm/$c$ the expansion collective
momentum keeps almost unchanged in the AMD calculations.  The mean
field restoring force, that would recompact the system, appears less
effective in the event-averaged one-body dynamics.  Moreover, after this
time, the radial dependence of the collective momentum is almost
self-similar for the `liquid' part.  At all the time, we see a kink in
the $v(r)$ curve.  The exterior and rapid component corresponds to the
pre-equilibrium particles and the inner and slower component
corresponds to the `liquid' composed mainly of fragments.  However 
the kink is always more pronounced in the BGBD
case, pointing towards a larger difference of velocity between
fragments and pre-equilibrium particles.

\subsection{Fluctuations}
The analysis of the density variance is an important tool to identify
the moment when fragmentation sets in and to get some information
about the related mechanism.  During the fragmentation process, the
density variance, evaluated at a given position ${\bf r}$,
grows in time, then it saturates (when fragments do
not interact anymore, apart from the Coulomb repulsion) and eventually
it decreases due to the fact that fragments fly apart from each other.
For instance, in the spinodal decomposition scenario,
the density fluctuations grow exponentially, 
with a
characteristic amplification time $\tau$
(typically of the order
of 30-50 fm/$c$) related to specific properties of the
nuclear interaction such as its range \cite{rep}.

Density variances have been calculated starting from the value of the
density along the $x$-, $y$- and $z$-axes.  For instance, the variance
of the density along the $z$-axis is defined by
$S^2_z(z)=\langle(\rho(0,0,z)-\langle\rho(0,0,z)\rangle)^2\rangle$,
where $\rho(x,y,z)$ is the density for each event and the brackets
$\langle\ \rangle$ stand for the average value over the event
ensemble.


\begin{figure}
\includegraphics[width=8.cm]{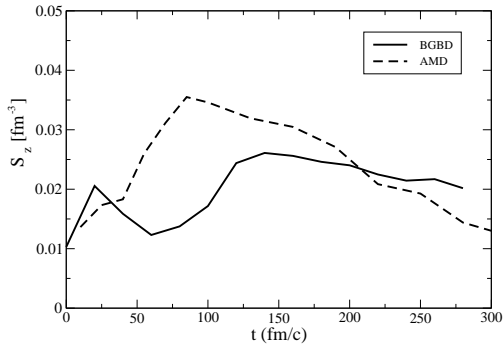}
\caption{Time evolution of the density variance, as obtained in BGBD (full line) and
AMD (dashed line) calculations.}
\label{fluct}
\end{figure} 

In Fig.\ \ref{fluct} we show the time evolution of $S_z$, which
is a representative value of $S_z(z)$, defined by
\begin{equation}
S_z=\frac{\int S_z(z)\langle\rho(0,0,z)\rangle dz}
         {\int \langle\rho(0,0,z)\rangle dz}.
\label{fluctav}
\end{equation}
Similar behaviour is observed for the variance in the other
directions, that can be defined analogously.  One can see that density
fluctuations have close values in the two approaches in the high
density phase ($t\approx 40$ fm/$c$). 
However, in the BGBD case
fluctuations are damped while the system expands, relaxing towards
the lower equilibrium value expected for nuclear matter
at lower density and temperature.
Only when low-density values are reached ($t\approx 70$ fm/$c$)  
and the
mean-field is unstable, one starts to see the rapid increase of
density fluctuations, corresponding to fragment formation. 
This is due to the fact that density
bumps are amplified, leading to fragments at normal density.  When
fragments fly apart from each other at large time $t$, the value of $S_z$
gradually decreases due to a trivial effect of the definition of $S_z$
[Eq.\ (\ref{fluctav})].

On the other hand, in the AMD case the density fluctuations continue
to grow as the system expands from $t\approx40$ to 70 fm/$c$,
suggesting that prefragments start to develop gradually at this stage
even though the density is not very low.  This is probably due to the
nucleon wave packet localization when two-body scattering occurs.  As
we have already mentioned, the relatively small magnitude of the
collective momentum observed in Fig.\ \ref{vel_AMD} around
$t\approx70$ fm/$c$ may be associated with the large momentum
fluctuation.  If the velocities of prefragments are randomly
distributed, the collective momentum will fluctuate from one event to
another.
\begin{figure}
\includegraphics[width=7.2cm]{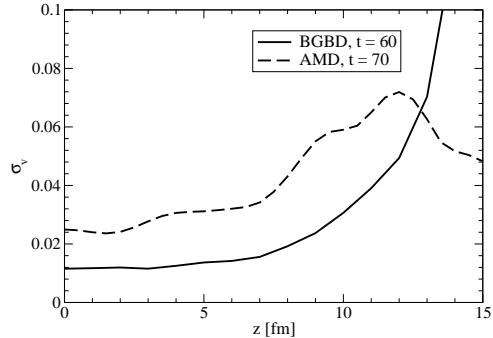}
\caption{Collective momentum variance, as obtained in BGBD (full line) and
AMD (dashed line) calculations, as a function of the distance on the z axis.}
\label{fluct_v}
\end{figure} 
This is confirmed by the analysis of collective momentum fluctuations,
displayed in Fig.\ \ref{fluct_v}.  In fact, when the system is expanding
($t = 60$ - 70 fm/$c$), larger fluctuations are seen in the AMD case.
Furthermore, the mean-field restoring force for the global expansion
dynamics is expected to be less effective in this situation 
and hence deceleration of expansion is weak in the AMD case as
observed in the comparison of Figs.\ \ref{vel_BGBD} and \ref{vel_AMD}.





\section{Discussion}

According to the results presented above, emission mechanisms appear
rather different in the two models.  For light particles, a more
abundant pre-equilibrium emission ($Z<3$) is observed in the BGBD
case.  Concerning the IMF emission mechanism, the involved time scales
and the relative importance of one- and many-body effects appear
different in the two models, due to the differences in the
implementation of mean-field propagation and two-body scattering.  In
the BGBD case the fragmentation process follows the spinodal
decomposition scenario.  Indeed the system enters the low density
region as a nearly homogeneous source, then density fluctuations are
amplified.  The formation of bubble-like structures is favoured, as a
manifestation of monopole instabilities.  In the AMD model, many-body
correlations have a stronger impact on the fragmentation dynamics,
while mean-field effects appear weaker.  Fragment formation seems not
to be directly linked with the spinodal decomposition scenario, but
rather with the appearance of some pre-fragments at earlier times.  As
a consequence, fragments are formed on shorter time-scales in AMD.
Finally, we observe that pre-equilibrium and IMF emissions happen on
different time scales and are related to different mechanisms in the
BGBD case, while in AMD the distinction of the two time scales is not
as clear as in BGBD.

As shown in Fig.\ \ref{fluct}, spatial density fluctuations have
similar values, in the high density phase, in the two models.
However, when the system starts to expand, fluctuations are quenched
in BGBD (until mean-field instabilities are encountered) while they
increase in AMD. The same is true for the collective momentum
fluctuations (see Fig.\ \ref{fluct_v}).  These differences are 
  naturally understood as an effect of the nucleon localization in the
AMD case, that induces larger many-body correlations. 
One can conclude that, in the energy range considered here (50 MeV/nucleon),
the fragmentation path is really sensitive to the interplay between
one- and many-body effects.  Changing the relative weight of these
effects leads to a rather different outcome.

As a consequence of the emission mechanisms outlined above, one expects
to see different primary fragment configurations in the two models,
with larger fluctuations in the AMD case,
though results could 
be similar for inclusive observables, such as charge distributions, especially
after secondary decay effects have been considered.
Considering kinematic properties and more exclusive observables,
such as IMF-IMF correlations, that keep a better track of the
freeze-out configuration and primary fragment partitions, one should
be able to disentangle between the predictions of the two models.  A
comparison with available experimental data would allow to shed some
light on the mechanisms and most relevant effects that govern the
fragmentation process.  This analysis will be the subject of a
forthcoming paper.

 Also, it would be interesting to extend this study 
to semi-peripheral
collisions, to investigate transport properties and dissipation-equilibration 
mechanims. In charge asymmetric reactions, isospin equilibration
could be a good tracer of the reaction dynamics. 




\begin{acknowledgments}
We thank M.Di Toro for the reading of the manuscript and stimulating discussions.

The AMD calculation was partly supported by the High Energy
Accelerator Research Organization (KEK) as a supercomputer project.
\end{acknowledgments}

\end{document}